\documentclass{aa}  

\usepackage{graphicx}
\usepackage{color}
\usepackage{txfonts}
\usepackage{lineno}
\begin{document} 


   \title{The Curie line in protoplanetary disks and the formation of Mercury-like planets}

  \author{T. Bogdan
        \inst{1}\thanks{E-mail: tabea.bogdan@uni-due.de}
        \and
        C. Pillich\inst{2}
        \and
        J. Landers\inst{2}
        \and
        H. Wende\inst{2}
        \and
        G. Wurm\inst{1}
}
\authorrunning{Bogdan et al.}

\institute{University of Duisburg-Essen, Faculty of Physics,
        Lotharstr. 1, 47057, Germany\\
        \and
        University of Duisburg-Essen and Center for Nanointegration Duisburg-Essen (CENIDE), Faculty of Physics, Lotharstr. 1, 47057 Duisburg, Germany\\
}

\titlerunning{The Curie line}
   \date{Received 30 September 2022 / Accepted 14 November 2022 / Published online 27 January 2023 }

 
  \abstract{
  In laboratory experiments, we heated chondritic material up to 1400\,K in a hydrogen atmosphere. M\"ossbauer spectroscopy and magnetometry reveal that, at high temperatures, metallic iron forms from silicates. The transition temperature is about 1200\,K after 1\,h of tempering, likely decreasing to about 1000\,K for longer tempering. This implies that in a region of high temperatures within protoplanetary disks, inward drifting solids will generally be a reservoir of metallic iron. Magnetic aggregation of iron-rich matter then  occurs within the magnetic field of the disk. However, the Curie temperature of iron, 1041\,K, is a rather sharp discriminator that separates the disk into a region of  strong magnetic interactions of ferromagnetic particles and a region of weak paramagnetic properties. We call this position in the disk the Curie line. Magnetic aggregation will be turned on and off here. On the outer, ferromagnetic side of the Curie line, large clusters of iron-rich particles grow and might be prone to streaming instabilities. To the inside of the Curie line, these clusters dissolve, but that generates a large number density that might also be beneficial for planetesimal formation by gravitational instability. One way or the other, the Curie line may define a preferred region for the formation of iron-rich bodies. }

   \maketitle
%

\section{Introduction}

Current observations of protoplanetary disks show a large suite of ring structures at various distances for solids and gas \citep{zhang2015evidence, Long2018, Andrews2018, Zhang2021, Miotello2022}. Just as large is the suite of possible explanations, ranging from embedded planets over pressure traps and temperature-dependent sticking properties to photophoretic concentration, snow lines, and local instabilities \citep{Krauss2005, Charnoz2021, Jiang2021, Pillich2021, Kuwahara2022, Lesur2022, Li2022}. 
We propose another ring mechanism here within the inner 1 AU that might be tied to the formation of Mercury-like planets. 

There has always been, and still is, a need to explain the high density of Mercury. This planet obviously has an iron core that is much larger than its small size would allow for if Mercury were composed of the same Solar System material as the other terrestrial planets \citep{Spohn2001}.
There have also been a number of exoplanets detected so far that might share Mercury's high density \citep{Rappaport2013, Sinukoff2017, Santerne2018, Lam2021}. 
This calls for mechanisms that can somehow separate iron from other material. This separation can occur before or after the planet formed. If it occurred afterward, part of the mantle would be lost one way or the other, that is, by one or more giant impacts or sublimation \citep{Cameron1985, Benz1988, Franco2022, Reinhardt2022}. 

If Mercury inherited its composition from the times of its formation, though, then all mechanisms that separate iron and silicates early on, before first planetesimals and planets form, might be relevant. Due to the different thermal conductivities, photophoresis  has been proposed to preferentially remove silicates but not the metal iron from the inner system \citep{Wurm2013}.
It has recently been proposed that evaporation and condensation trigger the formation of large iron-rich solids \citep{Aguichine2020, Johansen2022}. Also, super-Mercurys have been proposed to form if most of the iron is in pure form (Mah and Bitsch, personal communication). If that is true and metallic iron is already present, then ferromagnetism might also play a role. 

\citet{Nuth1994} and \citet{Nuebold2003} showed that a remanent magnetic field might already increase the capability of building larger aggregates with iron particles included. However, magnetic dipole interactions are much stronger if dipoles are aligned by an external magnetic field. Magnetic fields in protoplanetary disks exist, though it is far from straightforward to calculate or observe them \citep{Bertrang2017, Delage2022, Bjerkeli2016}. In any case, the estimated field strengths reach up to the millitesla range in the inner disk, decreasing with distance \citep{Wardle2007, Dudorov2014, Brauer2017}. With an increased stability of iron dipoles in mind, \citet{Hubbard2014} proposed an erosion mechanism that selectively erodes silicates while iron -- being bound better or being "stickier" in magnetic fields -- remains aggregated. While this has so far not been proven in experiments, \citet{Kruss2018} and \citet{Kruss2020} nevertheless showed that magnetization does allow larger clusters of iron-rich aggregates to grow where silicates have ceased to grow, but this occurs at the bouncing barrier and not during fragmentation. 

The bouncing barrier is the aggregate size of about 1\,mm to which micrometer dust can grow by hit-and-stick collisions until the aggregates become compact and can no longer dissipate enough collisional energy to grow further, that is, they only bounce off each other at this stage \citep{Zsom2010, Kelling2014}. However, with a significant fraction of metallic iron being present and with large enough magnetic fields, these aggregates can form larger chain-like clusters and, thus, continue to grow significantly.
In this way, \citet{Kruss2018} and \citet{Kruss2020} introduced the magnetic clustering of magnetic dust aggregates as a mechanism for selectively growing iron-rich bodies. 
In order for magnetic aggregation to work in protoplanetary disks, one requirement is the existence of metallic iron. 

\section{Sources of metallic iron}

Iron is present in meteorites in various oxidation states and absolute amounts. It is, for example, customary to plot the non-oxidized iron in metal and sulfides against the oxidized iron in silicates and oxides in a Urey-Craig diagram \citep{Weisberg2006, Krot2014}. This shows that ordinary chondrites are usually less oxidized than carbonaceous chondrites. This is due to the oxidation state in carbonaceous chondrites being strongly influenced by aqueous alteration \citep{Garenne2019}.
Ignoring CH and CB chondrites, where the iron content is exceptionally high, probably as a result of being impact melts, metallic iron still makes up as much as  10\,\% of chondritic meteorites by volume  \citep{Weisberg2006}. 

So metallic iron is present in protoplanetary disks. In general, metallic iron should have been formed at high temperatures, that is, as part of an (equilibrium) condensation sequence \citep{Ebel2006}. 
This places any processing of metallic iron in the inner, hot part of the protoplanetary disk. 
The details of iron condensation also depend on the composition of the disk and the ambient pressure \citep{Grossman2012}.
Fayalite, as an iron-bearing silicate, usually only forms up to 800\,K, with metal iron condensation occurring at higher temperatures. This might not explain the high fayalite content within chondrites \citep{Grossman2012}. However, \citet{Grossman2012} also conclude that supersaturated metallic iron as simulated by \cite{Fedkin2006},
which would solve the abundance problem with fayalite forming up to 1200\,K, would be unlikely in view of other condensables. Otherwise, FeS might also form from metallic iron at around 700\,K  \citep{Larimer1967}. This would allow metallic iron to be present at temperatures higher than 700\,K as, for instance, also suggested by recent calculations by \cite{Jorge2022}.
In any case, there might be a continuous supply of matter from the outside based on the classical "problem" that particles tend to drift inward \citep{Weidenschilling1977}. 
In fact, the inward drift of icy planetesimals was, for example, also considered in the context of providing a high water content to increase the oxygen fugacities in the hot inner region \citep{Ciesla2006}.
Therefore, there might be a continuous feed of further iron as solids drift inward. As the material drifts inward from the outside, it reaches regions of ever-increasing temperatures.  \citet{Ebel2011} consider condensation in a chondritic vapor atmosphere as a source of Mercury's composition. So the inward drift of chondritic matter might make sense here. 

To simulate the evolution of the sticking behavior upon inward drift, we measured the sticking force between the chondritic dust particles, the composition of which changed as the temperature increased. \citep{Bogdan2020, Pillich2021}. 
In the past we had used an ordinary L4/5 chondrite (Sayh al Uhaymir 001) for these studies \citep{Bogdan2020, Pillich2021} but have since extended theses studies to a CV chondrite (Allende).
This will not be discussed further here but will be reported elsewhere. However, as we heated parts of the "Allende" sample  up to 1400\,K in a hydrogen atmosphere, the iron was partially reduced to metallic iron at high temperatures, as tabulated in Table\,\ref{Moessbauer}.
The composition was  measured after the dust cooled to room temperature.

\begin{table}[]
    \centering
    \begin{tabular}{c|c}
        Temperature (K)& Fraction (\%) \\
          $<$ 1200 & 0.0 \\
          1200 & 4.6\\
          1250 & 8.7\\
          1300 & 13.9\\
          1350 & 18.6\\
          1400 & 25.8\\
    \end{tabular}
    \caption{Fraction by weight of the metallic iron phase from Mössbauer spectroscopy after heating Allende samples for 1\,h at a given temperature in a hydrogen atmosphere.}
    \label{Moessbauer}
\end{table}

The dust was only heated for 1\,h. It is likely that on larger timescales in protoplanetary disks more iron will more readily be produced at lower temperatures. Exemplary studies of heating a sample for 100\,h show that the threshold conditions for producing iron can be reduced by 200\,K or down to below 1000\,K. 

\subsection*{Magnetometry}
\begin{figure}
   \centering
   \includegraphics[scale=0.4]{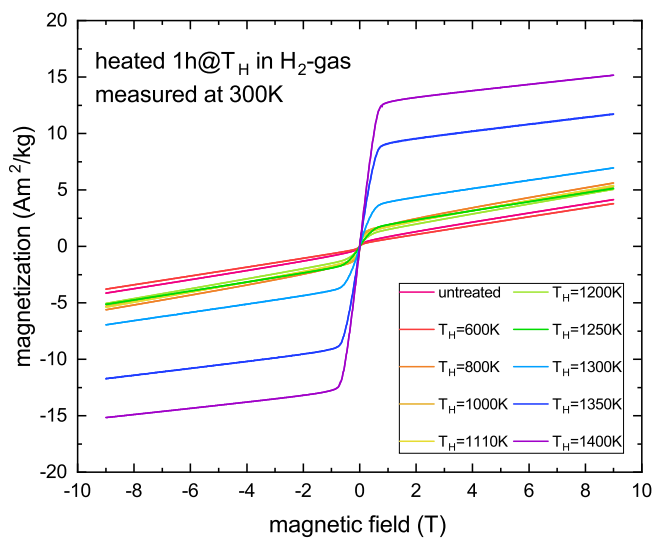}
      \caption{Magnetization over the applied magnetic field of the Allende powder samples for the different temperatures, $T_H$, at which it was tempered for 1\,h in a hydrogen atmosphere.
              }
         \label{Magnetometry_M(H)}
   \end{figure}

\begin{figure}
   \centering
   \includegraphics[scale=0.4]{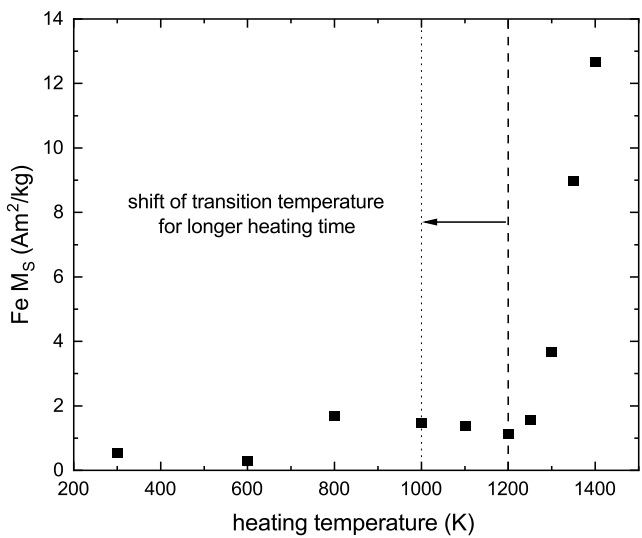}
      \caption{Fe saturation magnetization, $\mathrm{M_S}$, extracted from Fig.\,\ref{Magnetometry_M(H)}. The lines indicate the measured onset of metallic iron production after one hour of tempering and the potential shift estimated from a long-duration tempering experiment.
              }
         \label{Saturation_Magnetization}
   \end{figure}

The formation of metallic iron is additionally  traced by magnetometry. Figure\,\ref{Magnetometry_M(H)} shows the magnetization, $M$, for applied magnetic fields to demonstrate the effect of the newly formed phases on the magnetic properties. It should be noted that magnetic fields in protoplanetary disks are much smaller than the maximum fields applied here. 
Based on the shape of the M(H) curves and Mössbauer spectroscopy analysis, we assume the magnetization to consist of a ferromagnetic contribution of metallic iron, which already aligns at moderate fields, and a background of para- and diamagnetic contributions. The magnetization of the metallic iron was extracted via the interpolation of the high-field region to 0\,T.
The Fe magnetizations (Fig.\,\ref{Saturation_Magnetization}) clearly show a strong increase, by a factor of 2, between the untreated sample and the sample heated to 1000\,K, and a further increase, by a factor of 5, when heated to 1400\,K.

\section{Magnetic aggregation and the Curie line}

In view of the meteorite research and condensation calculations mentioned above, it might not come as a surprise that metallic iron forms at high temperatures within the meteorite sample in a reducing atmosphere. However, our experiments, which used a carbonaceous chondrite, confirm this with a mix of materials that is the closest composition to what might be expected in protoplanetary disks drifting inward.

If the iron is formed in protoplanetary disks, the material will have temperatures around the Curie temperature of 1041\,K for iron. We call the respective distance to the star the Curie line. Inside of the Curie line, the iron is paramagnetic, so it does not significantly react to the disk's magnetic field. Outside of the Curie line, however, the iron is ferromagnetic.
The spatial distances are not too large where phase changes occur and where the Curie line is situated. Therefore, with the inner region being a region of active magnetorotational instability turbulence, material can readily diffuse some distance \citep{Johansen2022}. In fact, if drift continued and all solids were sublimating, following \citet{Johansen2022}, selective nucleation of metallic iron might occur at about 1000\,K, just outside of the Curie line. If so, there would be pure iron or iron-rich particles both inside and outside of the Curie line. On the outside, the iron  grains would form larger clusters according to  \citet{Kruss2018, Kruss2020}. There are two options now for the fate of these iron-rich clusters. 

All the material continues to drift inward on average, as does the iron-rich matter until it hits the Curie line again. 
Option one is that these iron-rich clusters just drift inward and dissolve inside of the Curie line \citep{Kruss2018}. This might be in analogy to the snow line, where ice particles crossing the snow line evaporate and set dust grains free, creating a kind of traffic jam as the preferred site of planetesimal formation via direct gravitational instability \citep{Saito2011, Ida2016, Aly2021, Carrera2021}.
The other option would be that the growth of iron-rich clusters is so efficient that it directly enables concentrations of, for instance, streaming instabilities and iron-rich planetesimals forming just outside of the Curie line. This would also find its analogy on the outside of the snow line \citep{Hyodo2019, Ros2019}.

\begin{figure}
        \includegraphics[width=\columnwidth]{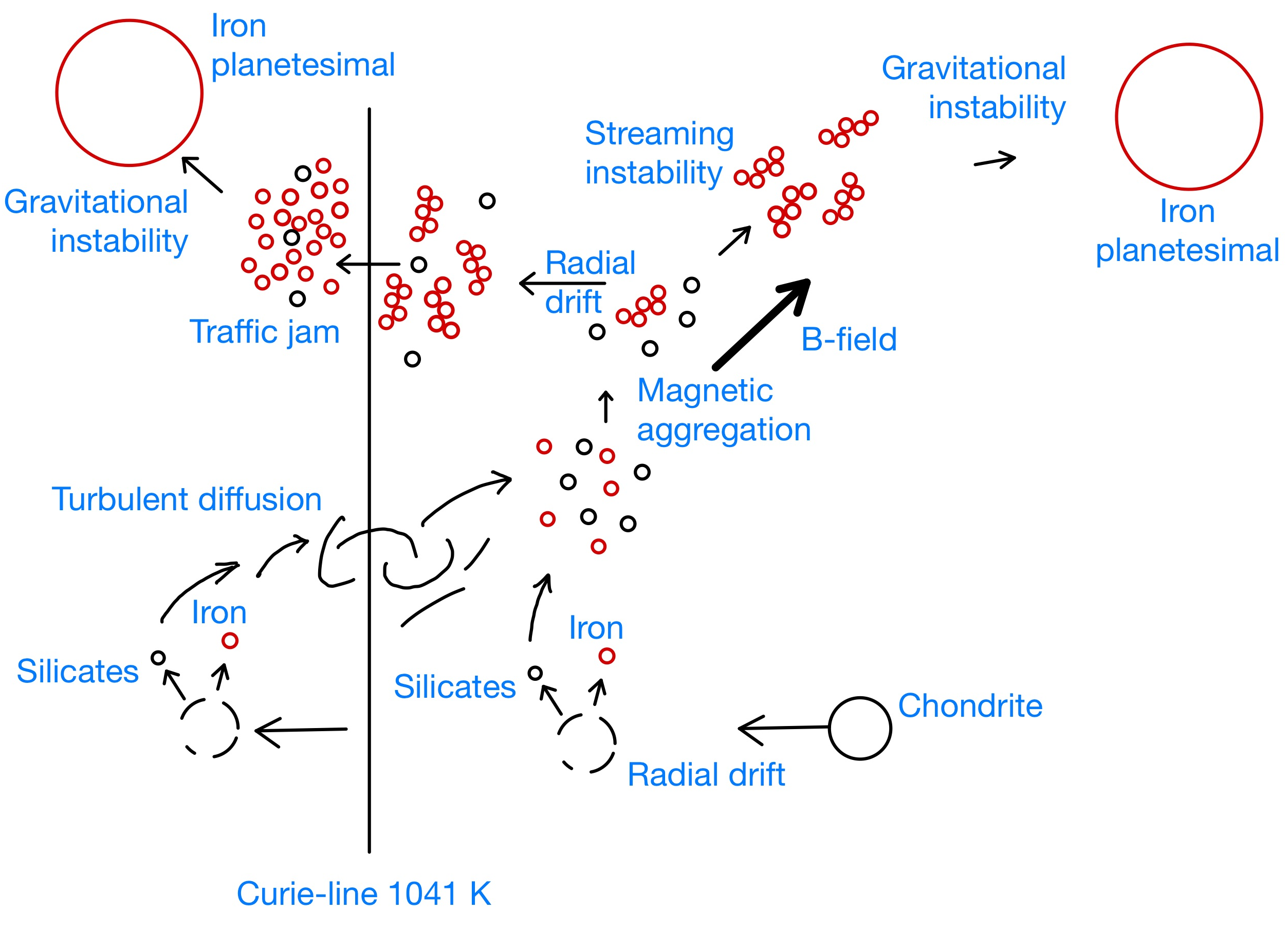}
        \caption{\label{howitworks} How the formation of iron planetesimals might work at the Curie line: chondritic matter drifts inward in protoplanetary disks. The iron is reduced to metallic iron at certain locations inside or outside of the Curie line. The Curie line separates magnetic aggregation from cluster dissolution. This provides two options for triggering iron-rich planetesimal formation. }
\end{figure}

Therefore, in any case, there will be a bias for iron-rich growth.\ What we propose here is that this might be a preferred line for iron-rich planetesimal formation in some analogy to the snow line farther outward. This is visualized in Fig.\,\ref{howitworks}.

\section{Conclusions}

In standard scenarios of protoplanetary disks, solid particles readily drift inward.
In the inner parts, at high temperatures of around 1000\,K or less, the iron will be reduced to metallic iron and iron-rich particles will be a significant part of the solid inventory. Incidentally, at 1041\,K, the Curie temperature of iron is just situated in this region. 
Therefore, there is a rather steep gradient at the Curie line that separates ferromagnetic from \mbox{paramagnetic} metallic iron. With disk magnetic fields being present, there will be large clusters of iron-rich particles on the outside and many smaller particles inside of the Curie line. As outlined in the Introduction, there are a number of recent works that advocate the formation of iron-rich bodies related to evaporation and condensation. The Curie line acts as a supporting mechanism here, as processing the matter at the Curie line might help trigger gravitational or streaming instabilities of iron-rich matter and therefore be a birthplace for Mercury-like planets.

\begin{acknowledgements}
      This work is supported by the
      \emph{Deut\-sche For\-schungs\-ge\-mein\-schaft, DFG\/} projects WE 2623/19-1 and WU 321/18-1.
\end{acknowledgements}

%
   \bibliographystyle{aa} 
   \bibliography{bib} 
%

\end{document}